\documentclass[12pt]{article}
\usepackage{graphicx}
\def\beq{\begin{equation}}
\def\eeq{\end{equation}}
\def\bea{\begin{eqnarray}}
\def\eea{\end{eqnarray}}
\def\nn{\nonumber}
\def\roughly#1{\mathrel{\raise.3ex\hbox
{$#1$\kern-.75em\lower1ex\hbox{$\sim$}}}}
\def\lsim{\roughly<}

\def\sss{\scriptscriptstyle}

\def\bd{B_d^0}
\def\bdbar{{\overline{B_d^0}}}

\def\ks{K_{\sss S}}

\def\ApNPqph{{\cal  A}^{\prime,q}  e^{i  \Phi'_q}}  \def\ApNPuph{{\cal
A}^{\prime,u}  e^{i  \Phi'_u}}  \def\ApNPdph{{\cal A}^{\prime,d}  e^{i
\Phi'_d}}   \def\ApNPCqph{{\cal   A}^{\prime   {\sss   C},   q}   e^{i
\Phi_q^{\prime C}}}  \def\ApNPCuph{{\cal A}^{\prime {\sss  C}, u} e^{i
\Phi_u^{\prime C}}}  \def\ApNPCdph{{\cal A}^{\prime {\sss  C}, d} e^{i
\Phi_d^{\prime C}}} \def\ApNPcomb{{\cal A}^{\prime, comb} e^{i \Phi'}}

\def\bra#1{\left\langle  #1\right|} \def\ket#1{\left| #1\right\rangle}
\def\barpk{{\raise.35ex\hbox  {${\sss  (}$}}--{\raise.35ex\hbox{${\sss
)}$}}}        \def\bbarp{\hbox{$B$\kern-0.9em\raise1.4ex\hbox{\barpk}}}

   \def\pewp{P'_{\sss  EW}}        
\def\ApNPqph{{\cal  A}^{\prime,q}  e^{i  \Phi'_q}}  \def\ApNPuph{{\cal
A}^{\prime,u}  e^{i  \Phi'_u}}  \def\ApNPdph{{\cal A}^{\prime,d}  e^{i
\Phi'_d}}   \def\ApNPCqph{{\cal   A}^{\prime   {\sss   C},   q}   e^{i
\Phi_q^{\prime C}}}  \def\ApNPCuph{{\cal A}^{\prime {\sss  C}, u} e^{i
\Phi_u^{\prime C}}}  \def\ApNPCdph{{\cal A}^{\prime {\sss  C}, d} e^{i
\Phi_d^{\prime C}}} \def\ApNPcomb{{\cal A}^{\prime, comb} e^{i \Phi'}}

\def\lsim{\roughly<}   \def\bd{B_d^0} 
\def\bdbar{{\bar   B}_d^0}   
    \def\ks{K_{\sss    S}}
           \def\pewp{P'_{\sss          EW}}
     
\def\btopik{B  \to   \pi  K}  \def\ApNPqph{{\cal   A}^{\prime,q}  e^{i
\Phi'_q}}    \def\ApNPuph{{\cal     A}^{\prime,u}    e^{i    \Phi'_u}}
\def\ApNPdph{{\cal  A}^{\prime,d}  e^{i \Phi'_d}}  \def\ApNPCqph{{\cal
A}^{\prime {\sss  C}, q} e^{i  \Phi_q^{\prime C}}} \def\ApNPCuph{{\cal
A}^{\prime {\sss  C}, u} e^{i  \Phi_u^{\prime C}}} \def\ApNPCdph{{\cal
A}^{\prime {\sss  C}, d} e^{i  \Phi_d^{\prime C}}} \def\ApNPcomb{{\cal
A}^{\prime, comb} e^{i \Phi'}}   
                  
\def\bra#1{\left\langle  #1\right|} \def\ket#1{\left| #1\right\rangle}

  \def\rr2{{1\over\sqrt{2}}}

\def\.{\!\cdot\!}    \def\:{\cdots}   \def\[{\left[}   \def\]{\right]}
\def\({\left(} \def\){\right)} 
\begin{document}

\begin{flushright} 
\end{flushright}

\begin{center}
\bigskip {\Large  \bf The 2-3  symmetry: Flavor Changing  $b$, $\tau$
Decays  and   Neutrino  Mixing\\}  \bigskip   {\large  Alakabha  Datta
$^{a}$\footnote{datta@physics.utoronto.ca}  and Patrick  J.  O'Donnell
$^{a}$\footnote{odonnell@physics.utoronto.ca} }
\end{center}

\begin{flushleft}
~~~~~~~~~~~$a$: {\it Department  of Physics, University of Toronto,}\\
~~~~~~~~~~~~~~~{\it  60 St.\  George Street,  Toronto, ON,  Canada M5S
1A7}\\
\end{flushleft}

\begin{center} 
\bigskip (\today) \vskip0.5cm {\Large Abstract\\} \vskip3truemm
\parbox[t]{\textwidth} {  The observed pattern of  neutrino mixing may
 be  the result  of a  2-3(  $ \mu-  \tau$) symmetry  in the  leptonic
 sector.  We consider a two Higgs doublet model with a 2-3 symmetry in
 the down type  quark and the charged lepton  sector.  The breaking of
 the 2-3 symmetry by the strange quark mass and the muon mass leads to
 FCNC  in the  quark sector  and the  charged lepton  sector  that are
 suppressed by ${  m_s \over m_b}$ and ${  m_{\mu} \over m_{\tau}}$ in
 addition to  the mass of  the heavy Higgs  boson of the  second Higgs
 doublet.  A Higgs  boson mass of $ m_H \sim 600 - 900$  GeV can explain the
 deviation   from  standard   model   reported  in   several  rare   B
 decays. Predictions for  other B decays are made and a  new CP phase is
 predicted  in  $B_{s}-{\bar{B}_{s}}$   mixing.   The  lepton  flavour
 violating  decays $  \tau  \to  \mu \bar{l}(\bar{q})  l(q)$  are below  the
 experimental  limits.  The  breaking of  2-3 symmetry  in  the lepton
 sector  can lead  to deviations  of the  atmospheric  neutrino mixing
 angle from the maximal value by $ \sim 2$ degrees.  }
\end{center}

\thispagestyle{empty} \newpage \setcounter{page}{1}
\baselineskip=14pt

\section{Introduction}

The  discovery  of  neutrino  masses  and  mixing  have  led  to  many
speculations about extension of the standard model(SM). In contrast to
the  mixing in  the quark  sector  which is  hierarchial, the  neutrino
mixing is large. A useful framework to understand the large neutrino mixing are
models    that     have    a    leptonic     $\mu-\tau$    interchange
symmetry\cite{mutau,moh}.    Several  ongoing   and   future  neutrino
experiments are expected to provide  us with insights into the physics
behind neutrino masses and mixing.

On the other  hand there is lot of data now  available on CP violation
in the  $B$ system. The  goal of  the B factories  is to test  the CKM
picture of CP violation and look for evidence of new physics.

{}  For some  time now  the B  factories have  been  reporting several
experimental hints of new physics. First,   within the SM, the measurement of
the CP  phase $ \sin {2 \beta}$  in $\bd(t) \to J/\psi  \ks$ should be
approximately  equal to that  in decays  dominated by  the quark-level
penguin transition $ b \to s q{\bar q}$ ($q=u,d,s$) like $\bd(t) \to \phi
K_s$,  $\bd(t) \to \eta^{\prime}  K_s$, $\bd(t)  \to \pi^0  K^0$, etc.
However, there is  a difference between the measurements  of $ \sin {2
\beta}$ in the $ b \to s $ penguin dominated modes ($ \sin {2 \beta} =
0.50 \pm 0.06$) and that in  $\bd(t) \to J/\psi \ks$($ \sin {2 \beta}=
0.685 \pm 0.032)$ \cite{ Babar, Belle,hfag}. Note that the
$ \sin {2 \beta}$ number for the $ b \to s $ penguin dominated modes
is the average of several modes.  The effect of new physics
can be different for different final hadronic states and so the
individual $ \sin {2 \beta}$ measurements for the different modes
are important.
Second, the  latest data on  $B\to\pi K$ decays (branching  ratios and
various  CP  asymmetries)  appear  to  be  inconsistent  with  the  SM
\cite{BKpidecays1,BKpidecays2a}.  {\footnote{ A cleaner test of the SM
could be provided by looking at the quasi-exclusive decays $B \to K X$
\cite{dattaKX} rather  than the  exclusive $B \to  K \pi$  decays.}  }
Third, within the SM, one  expects no triple-product asymmetries in $B
\to \phi K^*$  \cite{BVVTP}, but BaBar has measured  such an effect at
$1.7\sigma$ level  \cite{BaBarTP}.  There  are  also polarization  anomalies
where the decays $\bdbar \to \phi K^*$ and $B^- \to \rho^- K^*$ appear
to  have large  transverse  polarization amplitudes  in conflict  with
naive SM expectations \cite{hfag,phiKstarexp,rhoKstarexp}.

While  these  deviations certainly  do  not  unambiguously signal  new
physics(NP), they give reason to speculate about NP explanations of the
experimental data.  Furthermore, it is  far more compelling to find NP
scenarios that provide a single solution to all the deviations than to
look  for solutions  to  individual discrepancies.   Taking all  these
deviations seriously one is lead to certain structures of NP operators
that can  explain the present data\cite{rhoKstar}. The  question then is
what kind of NP models can generate these specific operator structures at
low energies.   One can be more  ambitious and ask if  such models may
have any connection with the physics behind neutrino mixing.  In fact,
the large $\nu_{\mu} -\nu_{\tau}$ mixing can lead to new effects in $b
\to  s$ transitions  through squark  mixing in  certain Supersymmetric
Grand   Unified  Theories   \cite{susygut}.   However,   the  dominant
contributions in this  scenario are new QCD penguin  amplitudes to $B$
decays involving  $b \to  s $ transitions and such contributions appear to be ruled
out by  the present $B \to K  \pi$
data\cite{BKpidecays1,BKpidecays2a}.

In this paper  we present a simple two Higgs doublet  model with a 2-3
interchange  symmetry in  the down  quark sector  like  the $\mu-\tau$
interchange symmetry in the leptonic  sector. This model can provide a
single solution to  the deviations from SM observed  in $B$ decays and
has interesting implications for  the leptonic sector.  It is believed
that one of the phenomenological problems of the 2-3 symmetry model is
the predictions $m_{\mu}=m_{\tau}$  or $m_{s}=m_{b}$. However, we will
assume the 2-3  symmetry in the gauge basis where  the mass matrix has
off  diagonal terms and is  2-3 symmetric.   Diagonalizing  the mass
matrix will split  the masses of $s$ and $b$ or  $\mu$ and $\tau$.  In
fact we  will consider the  ''enhanced" 2-3 symmetry where  the matrix
element of the mass matrix are invariant under any 2-3 interchange. In
other words we assume $<2|M|2>=<2|M|3>=<3|M|2>=<3|M|3>$. Diagonalizing the
mass matrix then  leads to
vanishing $m_s(m_\mu)$.

The breaking of the 2-3 symmetry is then introduced though the strange
quark  mass in  the quark  sector and  the muon  mass in  the leptonic
sector.  The  breaking of  the 2-3 symmetry  leads to  flavor changing
neutral currents  (FCNC) in  the quark sector  and the  charged lepton
sector that are suppressed by ${  m_s \over m_b}$ and ${ m_{\mu} \over
m_{\tau}}$ in addition to the mass  of the Higgs boson of
the second Higgs doublet. Additional  FCNC effects of similar size can
be generated  from the  breaking of the  $s-b$ symmetry in  the Yukawa
coupling of the second Higgs doublet.

In this paper we will be interested in the simplest two Higgs doublet  model with a 2-3
symmetric Yukawa coupling that can explain the hints of new physics observed in several rare B decays.
We, therefore, make some simplifying assumptions to 
make our model as predictive as possible. First,
we assume a discrete symmetry involving the down  quark 
to prevent FCNC effects in $ s \to d $ and $b \to d$ transitions. This allows us to 
satisfy constraints from measurements in the kaon system and $B_d$ mixing.
Second,
 we assume a simple ansatz for the
Yukawa coupling of the second Higgs doublet where the Yukawa matrix is described by
two real parameters and a universal weak phase. Finally, we assume there is no mixing
among the neutral Higgs bosons in the model and we only consider FCNC effects 
generated by the lightest Higgs bosons of the second Higgs doublet. Possible FCNC effects
from the other heavier neutral Higgs boson are neglected by assuming the mass 
of the Higgs boson to be sufficiently high. 

The  low energy  effective  Hamiltonian generated  by  this model  can
explain the  general features  of the deviations  from the SM  seen in
the various rare  B decays.  A new physics fit  to  the $  B  \to  K \pi$  data
\cite{BKpidecays2a}  allows us to fix the parameters of the model including  the lightest  Higgs mass  of the
second doublet. To extract the parameters of the model we have to calculate
the ratio of the tree amplitude in the standard model relative to a new physics amplitude.
We use factorization to calculate the tree amplitude and the
new physics amplitudes in the $K \pi$ system. Since non factorizable
effects in the tree and the new physics amplitudes are expected to be 
small \cite{BBNS,BKpidecays2b}
the use of factorization to extract the parameters of the model is reasonable.
 A more precise determination of the parameters of the model using the 
calculation of the
non leptonic amplitudes in the framework of QCD factorization\cite{BBNS}
will be carried out in a future work\cite{futurework}.

Having fixed the parameters of the model we make predictions   for the rare B decays $ B \to
\rho K^*, \phi K^*, \phi K_s$ and $\eta^{\prime} K_s$. Here again we  use factorization to
calculate the relevant non leptonic amplitudes. The inclusion of non factorizable effects
will change the predicted values of quantities such as
 polarization fractions, CP violation etc,
 but we do not expect the general pattern of new physics effects of the model 
in the various $B$ decays to change. This is because our predictions are largely dependent on
heavy quark theory considerations.
 More precise predictions for the various B decays in this model,
 including non factorizable effects,  will be carried out in a
future work\cite{futurework}.

We also make predictions
  for  $B_{d,s}-{\bar{B}_{d,s}}$  mixing  and for B  decays  with  the
underlying quark transitions $ b \to s \bar{u}u$ and $ b \to s \bar{c}
c$. The implication of the 2-3  symmetry in the up quark sector and on
the CKM  matrix is studied. This  model generates FCNC  effects in the
leptonic sector and  predictions for the decay $  \tau \to \mu \bar{l}
l$ and $ \tau \to \mu  \bar{q} q$ are made.  Finally, a deviation of about 2 degrees from
the maximal value for the atmospheric mixing  angle $ \theta_{23}$ is predicted from
the breaking of the 2-3 symmetry  in the charged lepton mass matrix.

We begin in Sec. 2 with a   
 description of the two Higgs doublet model with 2-3 interchange symmetry and
study the effects of the model in  various B decays. In 
Sec. 3 we study the implication of the model for the up quark sector 
and the CKM matrix. In Sec. 4 we study the effects of the model 
in the lepton sector, concentrating on  
 FCNC  effects in the $\tau$ decays and the neutrino mixing matrix. Finally in Sec. 5 we
conclude with a summary of the results reported in this work.

\section{ The Model- Quark Sector}

Two  Higgs  doublet models  (2HDM)  have  been  studied widely  and  a
particularly interesting  version with FCNC  was studied in  detail in
Ref.~\cite{soni}.  The model presented  here also has FCNC effects but
its origin and structure are different from that in Ref.~\cite{soni}.

We start with the discussion of our model in  the the quark sector.  Consider a Yukawa  Lagrangian of the
form  
\beq  {\cal  L}^{Q}_{Y}=  Y^{U}_{ij} \bar  Q_{i,L}  \tilde\phi_1
U_{j,R}  +  Y^D_{ij}  \bar  Q_{i,L}\phi_1 D_{j,R}  +  S^{U}_{ij}  \bar
Q_{i,L}\tilde\phi_2 U_{j,R} +S^D_{ij}\bar Q_{i,L} \phi_2 D_{j,R} \,+\,
h.c. ,
\label{lag1}
\eeq
\noindent where $\phi_i$, for $i=1,2$, are the two scalar doublets of
a 2HDM,  while  $Y^{U,D}_{ij}$ and $S_{ij}^{U,D}$  are the non-diagonal
matrices of the Yukawa couplings.

For convenience  we can choose to  express $\phi_1$ and  $\phi_2$ in a
suitable basis  such that  only the $Y_{ij}^{U,D}$  couplings generate
the fermion masses. In such a basis one can write,
  \beq 
  \langle\phi_1\rangle=\left(
\begin{array}[]{c}
0\\ {v/\sqrt{2}}
\end{array}
\right)\,\,\,\, , \,\,\,\, \langle\phi_2\rangle=0 \,\,\,.  \eeq
\noindent The two Higgs doublets in this case are of the form,
\bea   
\phi_1   &=  &   \frac{1}{\sqrt{2}}\pmatrix{0   \cr  v+H^0}   +
 \frac{1}{\sqrt{2}}\pmatrix{ \sqrt{2} \chi^+ \cr i \chi^0}, \nonumber\\
 \phi_2 &= & \frac{1}{\sqrt{2}}\pmatrix{ \sqrt{2} H^+ \cr H^1+i H^2}. \
 \eea
 
In principle there  can be mixing among the neutral  Higgs but here we
neglect such mixing.
We assume  the doublet $\phi_1$  corresponds to the scalar  doublet of
the SM and  $H^0$ to the SM Higgs field.  In  addition, we assume that
the second  Higgs doublet does  not couple to the  up-type quarks($S^U
\equiv 0$). For the down type couplings in Eq.~\ref{lag1} we have,
\beq   {\cal  L}^{D}_{Y}=  Y^D_{ij}   \bar  Q_{i,L}\phi_1   D_{j,R}  +
S^D_{ij}\bar Q_{i,L} \phi_2 D_{j,R} \,+\, h.c.
\label{lag2}
\eeq We  assume the  following symmetries for  the matrices  $Y^D$ and
$S^D$:

\begin{itemize}

\item{ There  is a discrete symmetry under  which $d_{L,R} \rightarrow
-d_{L,R}$}

\item{ There is a $s-b$ interchange symmetry: $s \leftrightarrow b$}

\end{itemize}

The discrete symmetry involving the  down quark is enforced to prevent
$ s \to d $ transition because of constraints from the kaon system. It
also prevents  $ b \to d $  transitions since $B_d$ mixing  as well as
the value  of $ \sin { 2  \beta}$ measured in $\bd(t)  \to J/\psi \ks$
are consistent with  SM predictions. Although there may  still be room
for NP  in $ b \to d$  transitions, almost all deviations  from the SM have
been reported only in $ b \to s$ transitions and so we assume no NP in
$ b \to d$ transitions in this work.

The above symmetries then give  the following structure for the Yukawa
matrices,
\begin{eqnarray}
 Y^D &= &\pmatrix{y_{11} & 0 & 0  \cr 0 & y_{22} & y_{23}\cr 0 &y_{23}
& y_{22}}, \nonumber\\ 
S^D &= &\pmatrix{s_{11} & 0 & 0 \cr 0 & s_{22} &
s_{23}\cr 0 &s_{23} & s_{22}}. \
 \label{Yukawas}
\end{eqnarray}

The down  type mass matrix,  $M^D$ is  now given by  $ M^D= {  v \over
\sqrt{2}}Y^D$.   The  matrix  $Y^D$  is  symmetric  and  choosing  the
elements in $Y^D$ to be real the mass matrix is diagonalized by,

\bea 
M^D_{diag} & = & U^T M^D U =\pmatrix{\frac{v}{\sqrt{2}}y_{11} & 0
&  0  \cr  0  &   \frac{v}{\sqrt{2}}(y_{22}-y_{23})  &  0\cr  0  &0  &
\frac{v}{\sqrt{2}}(y_{22}+y_{23})}, \nonumber\\ 
U &  = & \pmatrix{1 & 0
&   0  \cr   0   &  -\frac{1}{\sqrt{2}}   &  \frac{1}{\sqrt{2}}\cr   0
&\frac{1}{\sqrt{2}} &  \frac{1}{\sqrt{2}}}.\ 
\eea 
It is  clear that the
matrix $U$  will also diagonalize  the $S^D$ matrix when  we transform
the  quarks  from the  gauge  to the  mass  eigenstate  via $  d_{L,R}
\rightarrow U  d_{L,R}$. Hence there  are no FCNC effects involving  the Higgs
$\phi_2$.
 
The  down   quark  masses   are  given   by,
  \bea  m_d   &  =   &  \pm
\frac{v}{\sqrt{2}}y_{11},       \nonumber\\
  m_s       &=&      \pm
\frac{v}{\sqrt{2}}(y_{22}-y_{23}),   \nonumber\\
  m_b   &  =   &   \pm
\frac{v}{\sqrt{2}}(y_{22}+y_{23}).\ 
\eea

Since $m_s <<  m_b$ there has to be a  fine tuned cancellation between
$y_{22}$ and $y_{23}$ to produce  the strange quark mass. Hence, it is
more  natural to  consider  the symmetry  limit $y_{22}=y_{23}$  which
leads to  $m_s=0$. We then introduce  the strange quark  mass as a
small breaking of the $s-b$ symmetry and consider the structure,
\begin{eqnarray}
 Y^D_n &= &\pmatrix{y_{11} & 0 &  0 \cr 0 & y_{22}(1+2z) & y_{22}\cr 0
&y_{22} & y_{22}}, \
 \label{symbreak}
 \eea
with $z  \sim 2  m_s/m_b$ being a  small number.  Note that we  do not
break  the $s-b$ symmetry  in the  $2-3$ element  so that  the $Y^D_n$
matrix remains  symmetric. This down quark matrix  is now diagonalized
by,
\bea
M^D_{diag} & = & {W }^{T} M^D W 
 =\pmatrix{ \pm \frac{v}{\sqrt{2}}y_{11} & 0 &
0 \cr
0
&  \pm \frac{v}{\sqrt{2}}zy_{22} & 0\cr
0
&0 &
 \pm \frac{v}{\sqrt{2}}(2+z)y_{22}}, \nonumber\\
 W & = & 
 \pmatrix{1 & 0 &
0 \cr
0
& -\frac{1}{\sqrt{2}}(1-\frac{1}{2}z) & \frac{1}{\sqrt{2}}(1+\frac{1}{2}z)\cr
0
&\frac{1}{\sqrt{2}}(1+\frac{1}{2}z) &
 \frac{1}{\sqrt{2}}(1-\frac{1}{2}z)}.\
 \label{wmatrix}
 \eea
 In Eq.~\ref{wmatrix} we have dropped terms of $O(z^2)$.
 The down masses are now taken to be,
\bea
m_d & = &  \pm \frac{v}{\sqrt{2}}y_{11}, \nonumber\\
 m_s &=& \pm \frac{v}{\sqrt{2}}zy_{22}, \nonumber\\
m_b & = &  \pm \frac{v}{\sqrt{2}}(2+z)y_{22}.\
\eea
We find  $z \approx  \pm 2m_s/m_b$ and  now the transformation  to the
mass  eigenstate  will  generate   FCNC effects  involving  $  \phi_2$.  For
definiteness  we will  choose the  positive sign  for $z$  though both
signs are  allowed.  The matrix $S^D$  in the mass  eigenstate basis now has
the form,
\bea
S^D \rightarrow S^{D'} &= &\pmatrix{s_{11} & 0 &
0 \cr
0
& (s_{22}-s_{23}) & s_{23}z\cr
0
&s_{23}z &
 (s_{22}+s_{23}) }. \
 \label{sdm}
\end{eqnarray}
There  is now  FCNC  involving  a $  b  \to s  $  transition which  is
proportional to $z \sim m_s/m_b \sim \lambda^2$ where $\lambda$ is the
cosine of the Cabibbo angle.

It is also possible to find a $S^D$ that breaks the $s - b$ symmetry.
For example, we could choose,
\bea
 S^D &= &\pmatrix{s_{11} & 0 &
0 \cr
0
& s_{22} & s_{23}(1+ 2 \epsilon)\cr
0
&s_{23} &
 s_{33}}, \
 \label{sd1g}
 \eea
 with $ \epsilon $ being a small quantity of the same size or smaller
 than $z$.  A fit to the $ B \to K \pi $ data, presented later, will
 rule out small $z/\epsilon$.  The transformation to the mass
 eigenstate leads to the general parametrization,
\bea
S^D  \rightarrow  S^{D'} &= &\pmatrix{s_{d}e^{ i \phi_{dd}} & 0 &
0 \cr
0
& s_{s}e^{ i \phi_{ss}} & se^{ i \phi_{sb}} -p  e^{ i \psi_{sb}}\cr
0
&se^{ i \phi_{sb}}+p  e^{ i \psi_{sb}} &
 s_{b}e^{ i \phi_{bb}} }, \
 \label{sdm1general}
\end{eqnarray}
where
 \bea
s_{d}e^{ i \phi_{dd}} & = & s_{11}, \nonumber\\
s_{s}e^{ i \phi_{ss}} & =& (\frac{1}{2}s_{22}+ \frac{1}{2}s_{33}-s_{23})+
(\frac{1}{2}s_{33}-\frac{1}{2}s_{22})z-s_{23} \epsilon, \nonumber\\
s_{b}e^{ i \phi_{bb}} & = & (\frac{1}{2}s_{22}+ \frac{1}{2}s_{33}+s_{23})+
(\frac{1}{2}s_{22}-\frac{1}{2}s_{33})z + s_{23} \epsilon,   \nonumber\\ 
se^{ i \phi_{sb}} & = &(\frac{1}{2}s_{33}- \frac{1}{2}s_{22})+ 
s_{23}z, \nonumber\\
pe^{ i \psi_{sb}}& = & s_{23} \epsilon . \ 
\label{sd1m}
\eea
There is now an additional FCNC involving $ b \to s $ transitions
whose source is the $s-b$ symmetry breaking in $S^D$. Note that the
2--3 off--diagonal elements in $S^{D'}$  contain a part that is symmetric under
$s-b$ interchange and a part that is antisymmetric under the $s-b$
interchange.  The parameters in $ S^{D'}$ can be obtained or
constrained from a fit to $B$ decay data.

We will not consider this general case in the paper but to make our model predictive
 we will assume a simplified ansatz for
$S^D$ in Eq.~\ref{sd1g}. Here we will consider the following structure for the $S^D$
matrix with a small $s-b$ symmetry breaking,
\bea
 S^D &= &e^{i \phi}\pmatrix{
 s & 0 &
0 \cr
0
& 0 &  \pm s( 1+ 2\epsilon)\cr
0
& \pm s & 0
}, \
\label{sdsimple}
\eea
where $s$ and $\epsilon$ are real numbers and we have introduced a
universal weak phase $\phi$. 
On moving to
the mass eigenstate we obtain,
\bea
S^D \rightarrow S^{D'} &= & e^{ i \phi}\pmatrix{s & 0 &
0 \cr
0
& \mp s( 1+ \epsilon)  & \pm s (z- \epsilon)\cr
0
&\pm s (z+ \epsilon) &
 \pm s(1+\epsilon) }. \
 \label{sdmsimple}
\end{eqnarray}
Here the coupling  of the Higgs to  the $d$ and the $s$  quark are the
same, up to a sign, to  a good approximation since $\epsilon$ is a small
parameter.  The  FCNC $ b \to  s$ transition arises  from two sources.
The  first, represented  by  $z \sim  {  2 m_s/m_b}$,  comes from  the
breaking   of    the   $s-b$   symmetry   in    the   matrix   $Y^D_n$
(Eq.~\ref{symbreak})  and is symmetric  under $s-b$  interchange while
the  second  represented by  $  \epsilon \lsim  z$  ,  comes from  the
breaking   of   the   $s-b$    symmetry   in   the   matrix   $S^D$(
Eq.~\ref{sdsimple}) and is antisymmetric under $s-b$ interchange.

The size of  the matrix elements in $S^D$  are  not known.
One might  argue that the  size of the  matrix elements should  be the
same as the  matrix elements in the $Y$ matrix that  gives mass to the
down quarks. Hence the elements in $S^D$ are given as
 $ S^D \sim  Y \sim m_q/v$ where $q=d,s,b$ and $v \sim 246$ GeV
and are small.  However it is possible that the  down quark masses get
their  masses  from  a  different  Higgs  boson  than  the  top  quark
\cite{Das}, in which  case the vev of $\phi_1$ giving  mass to the down
quarks can  be small, much  less than $v$  , thus allowing  for larger
values of  $Y$ and $S^D$. Note that it is not necessary for
$Y$ and $S^D$ to be related and thus
  we assume  no relation between them.
 
 The lagrangian describing the interaction of the Higgs, $H_{1,2}$, is given by,
 \bea
 {\cal {L}}_1 & = & \frac{1}{2 \sqrt{2}} J (H_1+iH_2), \nonumber\\
 J_{ij}^{H_{1,2}} & = & S_{ij} \bar{d}_i(1+ \gamma_5)d_j  \pm S_{ij}^* \bar{d}_j(1- \gamma_5)d_i.\
 \label{lag}
 \eea
 Specifically  for $b \to s \bar{q}q$( $ q= d,s$) transitions the relevant currents, $J_{ij}$ are,
 \bea
 J_{sb}^{H_{1,2}}+J_{bs}^{H_{1,2}} & = & S_{sb} \bar{s}(1+ \gamma_5)b \pm 
 S_{bs}^* \bar{s}(1- \gamma_5)b,\nonumber\\
 J_{qq}^{H_{1,2}} & = & S_{qq} \bar{q}(1+ \gamma_5)q  \pm 
S_{qq}^* \bar{q}(1- \gamma_5)q.\
 \label{current}
 \eea
  Using Eq.~(\ref{sdmsimple}) we have,
  \bea
 S_{sb(bs)} & = &  \pm s e^{ i \phi}(z \mp \epsilon),\\
 S_{dd} & =& -S_{ss}  = \pm s  e^{ i \phi}. \
 \label{matrix}
 \eea

After integrating out the heavy Higgs boson, $H_{1,2}$,  we can generate the following effective
Hamiltonian with four quark operators,
\bea
H_{H_{1,2}}^{eff} &= & H_{eff}^s+H_{eff}^a, \nonumber\\
H_{eff}^s & = & \eta_{qq}\frac{G_F}{\sqrt{2}}\frac{2m_s}{m_b} \frac{s^2}{g^2}
\frac{m_W^2}{m_{H_{1,2}}^2}
\left[\pm O_{RR}  \pm O_{LL} +O_{RL} +O_{LR} \right], \nonumber\\
H_{eff}^a & = & \eta_{qq}\frac{G_F}{\sqrt{2}}\epsilon \frac{s^2}{g^2}
\frac{m_W^2}{m_{H_{1,2}}^2}
\left[\pm O_{LL}  \mp O_{RR} +O_{LR} -O_{RL} \right], \nonumber\\
O_{RR} & = & e^{ i 2 \phi} \bar{s} ( 1+\gamma_5) b \bar{q} (1+ \gamma_5) q, \nonumber\\
O_{LL} & = & e^{ -i 2 \phi} \bar{s} ( 1-\gamma_5) b \bar{q} (1- \gamma_5) q, \nonumber\\
O_{RL} & = &  \bar{s} ( 1+\gamma_5) b \bar{q} (1- \gamma_5) q, \nonumber\\
O_{LR} & = &  \bar{s} ( 1-\gamma_5) b \bar{q} (1+ \gamma_5) q, \
\label{heff}
\eea
where $\eta_{qq} =1(-1)$ for the choices in Eq.~(\ref{sdmsimple}), $g$
is the weak coupling and the plus (minus) signs in front of
$O_{LL,RR}$ correspond to the Higgs exchange $H_{1,2}$.  Note that a
new weak phase is associated only with the operators $O_{LL,RR}$ for
the ansatz in Eq.~\ref{sdmsimple}. The Higgs effective potential will
be assumed to make one of the neutral Higgs lighter than the other and
we will be interested in the phenomenology of the lightest Higgs only
which we will generically denote as $H$ and for the sake of
simplification and clarity we choose $H=H_1$ in Eq.~(\ref{heff}). Possible
FCNC effects associated with the other neutral Higgs boson will be neglected
by taking its mass to be sufficiently heavy. An
effective Hamiltonian will also be generated by the exchange of the
charged $H^+$ Higgs boson which will generate transitions of the form
$b \to c \bar{q} q$. We will not consider such effects in this work.

Now from the structure of $H_{H}^{eff}$ we can calculate nonleptonic
decays where hints of deviations from the SM have been reported.  We
will make a semi--quantitative analysis of NP effects in the various
decays and a more thorough investigation will be carried out in a
later work \cite{futurework}. 
We will first fix the parameters of the model from the fit results to the
$ K \pi$ system obtained in Ref.~\cite{BKpidecays2a}. We will then
make predictions for several nonleptonic $B$ decays.
For the purpose of the paper we will use
factorization to calculate nonleptonic decays. The  impact of
non factorizable effects will be addressed later in the paper.  For the
decay $B \to P_1 P_2$, where $P_{1,2}$ are any final state particles, the matrix element of any
operator in Eq.~(\ref{heff}), in factorization, has the structure $<O>\sim <P_1| \bar{s}
\gamma_A b|B><P_2| \bar{q} \gamma_B q|0>$ where $\gamma_{A,B} = ( 1
\pm \gamma_5)$ if the meson $P_2$ contains the $\bar{q}q$ component in
its flavour wavefunction. An example of such a decay is $ B^- \to K ^-
\pi^0$ with $ P_{1,2} \equiv K^-( \pi^0)$. On the other hand for the
decay $ B^- \to K^0 \pi^-$ the operators in Eq.~\ref{heff} have to be
given a Fierz transformation. We can define the Fierzed operators,
\bea
O_{LL}^F & = & -\frac{1}{2N_c}e^{ -i 2 \phi} \bar{q} ( 1-\gamma_5) b \bar{s} (1- \gamma_5) q 
-\frac{1}{8N_c}e^{ -i 2 \phi} \bar{q}\sigma_{\mu \nu} ( 1-\gamma_5) b \bar{s}\sigma^{\mu\nu} (1- \gamma_5) q, 
\nonumber\\
O_{RR}^F & = & -\frac{1}{2N_c}e^{ i 2 \phi} \bar{q} ( 1+\gamma_5) b \bar{s} (1+ \gamma_5) q 
-\frac{1}{8N_c}e^{ i 2 \phi} \bar{q}\sigma_{\mu\nu} ( 1+\gamma_5) b \bar{s}\sigma^{\mu \nu} (1+ \gamma_5) q, 
 \nonumber\\
O_{LR}^F & = &  -\frac{1}{2N_c}\bar{q}\gamma_{\mu} ( 1-\gamma_5) b 
\bar{s}\gamma^{\mu} (1+ \gamma_5) q, \nonumber\\
O_{RL}^F & = &  -\frac{1}{2N_c}\bar{q}\gamma_{\mu} ( 1+\gamma_5) b 
\bar{s}\gamma^{\mu} (1- \gamma_5) q, \
\label{fierz}
\eea
where we have done also a color Fierz and dropped octet operators that do not contribute in factorization.

We can now look at the various nonleptonic B decays and we start with
those decays with the underlying quark transition
$ b \to s \bar{d} d$.

\begin{itemize}

\item{ {\bf $B \to  K  \pi$ decays}:  
    
Let us denote by $A^{i j}$ the amplitude for the decay $\bd \to
\pi^i K^j$. In the SM they are described to a good approximation
by a ``tree'' amplitude $T'$, a gluonic ``penguin'' amplitude or
$P'$, and a colour-favored electroweak penguin (EWP) amplitude
$\pewp$.  In $B\to\pi K$ decays, there are four classes of NP
operators, differing in their color structure: ${\bar s}_\alpha
\Gamma_i b_\alpha \, {\bar q}_\beta \Gamma_j q_\beta$ and ${\bar
s}_\alpha \Gamma_i b_\beta \, {\bar q}_\beta \Gamma_j q_\alpha$
($q=u,d$). The matrix elements of these operators are then
combined into single NP amplitudes, denoted by $\ApNPqph$ and
$\ApNPCqph$ respectively with $q=u,d$.  In the presence of NP the
most general $\btopik$ amplitudes take the form
\cite{BKpidecays2a,BKpidecays2b},
\bea
\label{BpiKNPamps}
A^{+0} &\!\!=\!\!& -P' + \ApNPCdph ~, \nn\\
\sqrt{2} A^{0+} &\!\!=\!\!& P' - T' \, e^{i\gamma} - \pewp
+ \ApNPcomb - \ApNPCuph ~, \nn\\
A^{-+} &\!\!=\!\!& P' - T' \, e^{i\gamma} - \ApNPCuph ~, \nn\\
\sqrt{2} A^{00} &\!\!=\!\!& -P' - \pewp + \ApNPcomb +
\ApNPCdph ~, 
\eea
where $\ApNPcomb \equiv - \ApNPuph + \ApNPdph$, ${\cal A}^{\prime {\sss C},u}$,
 and ${\cal A}^{\prime {\sss C}, d}$ are the NP amplitudes 
\cite{BKpidecays2b}.

It was found in Ref.~\cite{BKpidecays2a}  
that a good fit to the data
can be obtained with the NP amplitudes
 $| {\cal A}^{\prime, comb} / T' | = 1.64$, $\Phi' \sim 100^0$,
 ${\cal A}^{\prime {\sss C},u} \approx 0$,
  and
 ${\cal A}^{\prime {\sss C}, d} \approx 0$. 
It is clear from the the structure of Eq.~\ref{heff} that
${\cal A}^{\prime {\sss C},
u} = 0$ as  the NP involves only down type quarks. 

Now one obtains, using Eq.~\ref{heff} and factorization, the matrix element 
relations for the decay $B^- \to K^- \pi^0$,
\bea
<O_{LL}(O_{LL}^F)> & = & -<O_{RR}(O_{RR}^F)> \nonumber\\
<O_{LR}(O_{LR}^F)> & = & -<O_{RL}(O_{RL}^F)>. \
\label{pp}
\eea
These relations follow from the fact that in factorization,
\begin{eqnarray*}
<O_{AB}> & \sim & <K^-| \bar{s}
\gamma_A b|B><\pi^0| \bar{d} \gamma_B d|0>,\
\end{eqnarray*}
 where $\gamma_{A,B} = ( 1
\pm \gamma_5)$. We then have,
\begin{eqnarray*}
<O_{LL}> & = & 
<K^-| \bar{s}(1-\gamma_5) b|B><\pi^0| \bar{d} (1-\gamma_5) d|0>,\nonumber\\
&=&-<K^-| \bar{s}b|B><\pi^0| \bar{d} \gamma_5 d|0>, \nonumber\\
&=&-<K^-| \bar{s}(1+\gamma_5) b|B><\pi^0| \bar{d} (1+\gamma_5) d|0>, \nonumber\\
&=&-<O_{RR}>.\
\end{eqnarray*}
 Similar arguments lead to the other relations in Eq.~\ref{pp}.
Using the matrix element relations in Eq.~\ref{pp} one obtains,  
\bea
<K^- \pi^0|H^{eff}_H|B^-> & = & |{\cal A}^{\prime, comb}|e^{ i\Phi'}= 
 \frac{G_F}{\sqrt{2}} A_{dd} 
\left[2 i\sin{2 \phi}\chi_s
-2 \cos{2 \phi}\chi_a + 2 \chi_a \right], 
\nonumber\\
\chi_{s} & = & \frac{2m_s}{m_b} \frac{s^2}{g^2}
\frac{m_W^2}{m_{H}^2},\nonumber\\
\chi_{a} & = & \epsilon\frac{s^2}{g^2}
 \frac{m_W^2}{m_{H}^2},\nonumber\\
A_{dd} & = & <K^-| \bar{s} b|B^-><\pi^0| \bar{d} \gamma_5 d|0>.\
\label{acomb}
\eea
To make estimates we choose $ \epsilon \sim z= {2m_s/m_b}$ and $\cos{2
  \phi} \sim - \sin {2 \phi}$. Choosing $2\phi=-45^0$ and the fact
that $\tan{ \Phi'} \approx \cot{ \phi}$ leads to $\Phi' \sim 113^0$,
which is consistent with the fit obtained in Ref.~\cite{BKpidecays2a}.
We observe that if $\chi_a/\chi_s=\epsilon/z \sim 0$ then $\Phi' \sim
90^0$ which is also consistent with the fit obtained in
Ref.~\cite{BKpidecays2a}. However, with $\chi_s/\chi_a=z/\epsilon \sim
0$ we obtain $\Phi' \sim 0$ which is inconsistent with the fit in
Ref.~\cite{BKpidecays2a}.  Hence only $\epsilon \lsim z$ is allowed as
indicated earlier.  Turning now to the amplitude ${\cal A}^{\prime
  {\sss C}, d}$ we have,
\bea 
<K^0 \pi^-|H^{eff}|B^-> & = & |{\cal A}^{\prime {\sss C}, d}|e^{\ i \Phi'_u},
\nonumber\\
&=& -\frac{1}{2N_c}
\frac{G_F}{\sqrt{2}}  
\left[2 F_{dd}( i \sin{2 \phi}\chi_s 
- \cos{2 \phi}\chi_a) + 2 \chi_a G_{dd} \right], \nonumber\\
F_{dd} & = & <\pi^-| \bar{d} b|B^-><K^0| \bar{s} \gamma_5 d|0>,\nonumber\\
G_{dd} & = & <\pi^-| \bar{d}\gamma_{\mu} b|B^-><K^0| \bar{s}\gamma^{\mu} \gamma_5 d|0>.\
\label{au}
\eea
It is clear from Eq.~(\ref{au}) that ${\cal A}^{\prime {\sss C}, d}$
is suppressed relative to ${\cal A}^{\prime {\sss C}, comb}$ by $ { 1
  \over ( 2 N_c)}$, and hence small, which is again consistent with
the fit obtained in Ref.~\cite{BKpidecays2a}.

Now using $| {\cal A}^{\prime, comb} / T' | = 1.64$ we find, using
naive factorization\cite{rhoKstar},
\bea
1.64 & = & \left \vert \frac{2k \sin{2 \phi}\chi_{s}A_{dd}}
{V_{ub}^*V_{us} \bra{\pi^0} {\bar b} \gamma_\mu (1 - \gamma_5) u \ket{B^-}
\bra{K^-} {\bar u} \gamma^\mu (1 - \gamma_5) s \ket{0} }
\right\vert,     \nonumber\\
k & = & \sqrt{1+\tan^2{ \phi}}.\
\label{constraint}
\eea
One can then convert this to
\beq
 2k |\sin{2\phi}|\chi_{s}  = { f_{\sss K} (m_{\sss B}^2 -
m_\pi^2) F_0^\pi / \sqrt{2} \over [ ( m_{\sss B}^2 - m_{\sss K}^2) /
(m_b - m_s) ] F_0^{\sss K} ( m_\pi^2 / 2 m_d ) f_\pi / \sqrt{2} } \,
1.64 \left( c_1 + {c_2 \over N_c} \right) \left\vert V_{ub}^* V_{us}
\right\vert ~. 
\eeq
We take $(f_{\sss K}/f_\pi) (F_0^\pi/F_0^{\sss K}) \sim 1$, $|V_{ub}^*
V_{us}/V_{tb}^* V_{ts}| = 1/48$ and $c_1 + c_2/N_c = 1.018$. Taking
the masses from the Particle Data Group \cite{pdg}, we find,
\bea
\chi_{s} & = & \frac{1}{k |\sin{2\phi}|}0.05 \frac{m_d}{6 MeV}|V_{tb}^* V_{ts}|.\
\label{chi}
\eea
This then leads to,
\bea
 \frac{m_W^2}{m_{H}^2} & = & \frac{g^2}{2 s^2 k|\sin{2 \phi}|}\frac{m_b}{m_s} 
\cases{ 0.033 |V_{tb}^* V_{ts}| &
$m_d = 4$ MeV, \cr 0.07 |V_{tb}^* V_{ts}| &
$m_d = 8$ MeV, \cr} \
\eea
and finally to, 
\bea
 m_W \frac{\sqrt{2}s}{g }\sqrt{\frac{m_s}{m_b }
 \frac{k|\sin{ 2 \phi}|}{0.07 |V_{tb}^* V_{ts}|}} & \le & m_{H} \le  m_W \frac{\sqrt{2}s}{g }\sqrt{\frac{m_s}{m_b }
 \frac{k|\sin{ 2 \phi}|}{0.033 |V_{tb}^* V_{ts}|}}, \nonumber\\
  4.7 M_W (\frac{s}{g}) & \le & m_H \le  6.8 M_W (\frac{s}{g}), \
\eea
where we have used $m_s=100$ MeV and $m_b=5$ GeV. For $s \sim 1$ we obtain
$m_{H} \sim 600- 900$ GeV.

Our estimate of the parameters of the model, like the new physics weak
phase $\phi$ and the higgs mass $m_H$,  used the factorization
assumption to calculate the nonleptonic amplitudes. It is therefore
important to address the impact of non factorizable effects on the
values of the  parameters of the model. We note that
to extract the parameters of the model we have to calculate
the ratio of the tree amplitude in the standard model relative to a new physics amplitude.
 Since non factorizable
effects in the tree and the new physics amplitudes are expected to be 
small \cite{BBNS,BKpidecays2b} we do not expect these 
effects to significantly alter the values of the parameters of the
model extracted using factorization. A more precise determination of
the model parameters using QCD factorization \cite{BBNS} will be carried out in a
future work \cite{futurework}.
}

\item{{\bf $B \to  K^*  \rho$ decays}:
 
Here the final states contain two vector mesons and so from angular momentum
    considerations there are three amplitudes. They are usually chosen
    to be  the longitudinal amplitude,  $A_0$, and the  two transverse
    amplitudes  $A_{\|}$  and  $A_{\perp}$. In the SM, $A_0$ is the dominant
    amplitude and the transverse amplitudes are $O(1/m_B)$ because of the
    $V-A$ structure of the weak interactions. Now   it  is  clear  from  the
    structure  of the  operators in  Eq.~(\ref{heff}) that  the scalar
    operators  do  not contribute  to  vector-vector  final states  in
    factorization.  Hence  we expect no new physics  effect in the decay  $B^- \to
    K^{-*} \rho^0$.  On  the other hand  the  decay $B^- \to K^{0*}
    \rho^{-}$ gets contribution from the the Fierzed operators in Eq.~\ref{fierz}.
    The tensor operators in $O_{LL,RR}^F$ contribute to the transverse
    amplitude  in   the  leading  order  in  the   large  $m_B$  limit
    \cite{rhoKstar} unlike in the SM.  This follows from the fact that
    the matrix  elements of  the tensor operators  contain the  piece $
    \sim    <K^{*0}(q_{\mu})|\bar{s}   \sigma_{\mu\nu}d|0>    \sim   (
    \epsilon_{\mu}q_{\nu}-\epsilon_{\nu}q_{\mu})$ that contribute to the
    transverse amplitudes in the large $m_B$ limit. {}For     a
    longitudinally polarized state  the polarization vector can be
    approximated as, $ \epsilon_{ \mu}
    \sim  q_{\mu}$ neglecting  $({m_{K^*}\over  m_B})^2$ effects,  and
    hence   the  contribution   to  the   longitudinal   amplitude, $A_0$, arise only  at
    $O(m_V/m_B)$.   The  operators  $O_{LR,RL}^F$  contribute  in  the
    leading  order  in  the  large  $m_B$ limit  to  the  longitudinal
    amplitude, $A_0$ while the transverse amplitudes are suppressed by
    $0({m_V \over m_B})$ \cite{BVVTP,rhoKstar}. 

Hence our prediction
    is that the decay   $B^- \to    K^{-*} \rho^0$ should be
    dominantly longitudinally polarized as in the SM because there is no new physics effect in this decay,
    while the  decay $B^- \to K^{0*}\rho^{-}$ may have a sizeable transverse polarization. These
    predictions appear to be consistent with present experiments where
   BaBar measures the longitudinal polarization fraction, $f_L$, for
$B^- \to    K^{-*} \rho^0$ to be
    $0.96^{+0.04}_{-0.15} \pm 0.04 $  while the
    BaBar and Belle measurements of $f_L$ for
 $B^- \to K^{0*}\rho^{-}$ 
are $0.79 \pm 0.08 \pm 0.04$ and $0.43 \pm 0.11^{0.05}_{-0.02}$
respectively giving an average $f_L$ of
    $0.66 \pm 0.07$, thereby showing sizeable transverse polarization in this mode
    \cite{rhoKstarexp, rhoKstar}. 

To calculate specific quantities we again use factorization to compute
the non leptonic amplitudes. We do not expect non factorizable effects
to significantly alter the pattern of new physics effects in the B
decays as our conclusions to a large extent rely on heavy 
quark theory considerations. As far as specific values of various
quantities, like polarization fraction, CP violation etc,
 in the B decays are concerned it is possible that non factorizable
 effects will change the values of these quantities. Hence,
our calculation of these quantities should be taken as rough estimates.
A more precise estimate of the quantities will be carried out in a
future work\cite{futurework}.

{} For the $A_{\perp}$ amplitude the factorized matrix element satisfy 
$<O_{LL}^F>=<O_{RR}^F>$ 
while for the $A_{\|}$ amplitude 
the factorized matrix element satisfy
$<O_{LL}^F>=-<O_{RR}^F>$  
and so one obtains,
\bea
A_{\perp} \sim 
\left[\chi_s \cos{2 \phi}-i \chi_a \sin{2 \phi}\right],\nonumber\\ 
A_{\|} \sim 
\left[-\chi_a \cos{2 \phi}+i \chi_s \sin{2 \phi}\right].\
\label{trans}
\eea  
Note that even though the Fierzed tensor operators are colour
suppressed their effects for the transverse amplitudes can nonetheless
be significant as the SM penguin amplitude for VV(V=vector) final
state is smaller than the PP (P=pseudoscalar) final states
\cite{rhoKstar}. It follows from Eq.~(\ref{trans}), for $\chi_s \sim
\chi_a$, that the transverse polarization fractions $f_{\perp} \sim
f_{\|}$ where $f_T \sim |A_T|^2$ ($T= \perp, \|$). We also expect no
triple product asymmetry \cite{BVVTP} from the interference of
$A_{\perp}$ and $A_{\|}$ because the relative weak phase difference is zero (or
$\pi$).
A triple product  asymmetry from the interference of the longitudinal amplitude $A_0$ and $A_{\perp}$
can be significant.


}
\end{itemize}

We now consider nonleptonic B  decays involving the underlying
 $ b \to s \bar{s} s$ transitions.
\begin{itemize}

\item{ {\bf $B \to  \phi  K_s$, $B \to  \eta^{\prime}  K_s$ decays}:
We begin with the decay $B \to  \phi  K_s$ where
only the Fierzed operators in Eq.~\ref{fierz} contribute 
 as $ \phi$  is a vector meson.  
In factorization, for this mode, we have the matrix elements
relations ,$<O_{LL}>=<O_{RR}>$  and $<O_{LR}>=<O_{RL}>$. 
 We can therefore write,
 \bea
 A(B_d \to \phi K_s) &= & A^{SM}+A^{NP}_{LR \pm RL} + A^{NP}_{LL \pm RR},\nonumber\\
 A^{SM} & = &-{G_{\sss F} \over \sqrt{2}}
 V_{tb}V_{ts}^* Z \left[ a_3^t+ a_4^t + a_5^t -\frac{1}{2}a_7^t
 -\frac{1}{2}a_9^t -\frac{1}{2}a_{10}^t \right. \nonumber\\ 
& & \hskip20truemm \left.  -a_3^c- a_4^c - a_5^c +\frac{1}{2}a_7^c
+\frac{1}{2}a_9^c +\frac{1}{2}a_{10}^c \right], \nonumber\\
A^{NP}_{LR \pm RL} & = &
-\frac{1}{N_c}\frac{G_F}{\sqrt{2}}Z \chi_{s},
\nonumber\\ 
A^{NP}_{LL \pm RR} & = & -\frac{1}{N_c}\frac{m_{\phi}}{m_B}\frac{G_F}{\sqrt{2}}Z 
\left[\chi_{s}  \cos{2\phi}  
-i\chi_{a} \sin{2\phi}\right], 
\nonumber\\
\chi_{s} & \sim & \chi_a=
 -\frac{1}{ k|\sin{2\phi}|}0.05 |V_{tb}^* V_{ts}|,\nonumber\\
Z &= & 2f_{\phi}m_{\phi}F_{BK}(m_\phi^2) \varepsilon^*\cdot p_B, \
\label{phiK}
\eea
where the SM contribution can be found in Ref.~\cite{dattarparity} and  we have chosen $m_d=6 $ MeV.

We can now calculate $\sin(2 \beta)_{eff}$ 
from,
\bea
\sin(2 \beta)_{eff} & = &-{2Im[\lambda_f] \over (1+|\lambda_f|^2)},\nonumber\\
\lambda_f & = &  e^{-2i\beta}{\bar{A} \over A},\
\eea
where $A=A_{\sss \phi K_S}^{\sss SM, }+ A_{\sss \phi K_S}^{NP}$ and
$\bar{A}$ is the amplitude for the CP-conjugate process.  Note that
from Eq.~(\ref{phiK}), $\sin(2 \beta)_{eff}$ is independent of $Z$ and
hence free from uncertainties in the form factor and decay constants.
We also observe that the deviation from the SM expectation of $\sin(2
\beta)_{eff}$ comes from $A^{NP}_{LL \pm RR}$ which is however colour
and $\frac{m_{\phi}}{m_B}$ suppressed, leading to a rough estimate,
$\sin(2 \beta)_{eff}\approx$ 0.43 which  agrees well with the present
experimental average 
of $0.44 \pm 0.27 \pm 0.05$
\cite{Babar,Belle,hfag} . Note that the
fact $A^{NP}_{LR \pm RL}$ does not carry any new weak phase is a
consequence of the form in Eq.~\ref{sdmsimple} and in the general case
( Eq.~\ref{sdm1general}) there will be a new physics phase associated
with this operator. Furthermore, if $\chi_a/\chi_s=\epsilon/z \sim 0$
then there is no new weak phase in this decay and the prediction of
$\sin(2 \beta)_{eff}$ should be given by the SM.

The decay $B \to \eta^{\prime} K_s$ is more complicated as both the $
b \to s \bar{d} d$ and $ b \to s \bar{s} s$ transitions contribute
\cite{dattaeta}. The final states are pseudoscalars and so both the
operators in Eq.~\ref{heff} and the Fierzed operators in
Eq.~\ref{fierz} for $ b \to s \bar{s} s$ transition contribute. Hence
in this model this decay  could show a significant deviation
from the SM prediction for $\sin(2 \beta)_{eff}$. For the "standard"
$\eta^{\prime}$ wave function $\eta^{\prime} \sim {( \bar{u} u +
  \bar{d}d- 2 \bar{s} s)\over \sqrt{6}}$ the NP $ b \to s \bar{d} d$
and $ b \to s \bar{s} s$ amplitudes add or partially cancel depending on the
choice in Eq.~\ref{sdmsimple}. In the $\eta$ they would cancel or add
producing a much smaller or bigger deviation from the SM in $B \to \eta
K_s$ compared to $B \to \eta^{\prime} K_s$. It should be noted that
the decay $B \to \eta K_s$ has a much smaller branching ratio,
probably because of destructive interference in the SM amplitudes
\cite{lipkin}.  }

\item{{\bf $\bdbar \to \phi K^*$ decays}: Here the final states
    contain two vector mesons like the $B^- \to \rho^- K^*$ decays and the
    predictions here are very similar.  This decay gets contributions only
    from the Fierzed operators in Eq.~\ref{fierz}. The tensor
    operators in $O_{LL,RR}^F$ can contribute significantly to the
    transverse amplitude in the large $m_B$ limit \cite{rhoKstar} thus
explaining the low longitudinal polarization fraction, $f_L=0.45 \pm
    0.05 \pm 0.02$ \cite{phiKstarexp,hfag}, measured in this decay.
          Given the large NP weak phase we could also observe a
    sizeable triple product asymmetry and/or direct CP asymmetry in
    this decay.  }

\end{itemize}
   
To summarize, the model can provide explanations for the deviation
from SM seen in several rare B decays.  It also makes specific
predictions in other B decays. For example, decays going through $ b
\to s \bar{u} u$ transitions like $ B_s \to K^+ K^-$ should  not be
affected.  The decays going through $ b \to s \bar{c} c$ transitions
like $ B_{d,s} \to D_s^{(+*)} D_{d(s)}^{(-*)}$ should also not be affected
and so this decay along with $ B_d \to D^{(+*)} D^{(-*)}$ can be used
to measure the angle $\gamma$ \cite{gammaDD} without NP pollution. The
model also has no effect in the decays $ B_{d,s} \to J/\psi
K^{(*)}(\phi, \eta')$.  A detail study of the predictions of this
model in B decays will be presented elsewhere \cite{futurework}.

We now discuss the important case of $B_{d,s}-{\bar{B}_{d,s}}$ mixing.
By construction of the model there is no effect in
$B_{d}-{\bar{B}_{d}}$ mixing.  In $B_{s}-{\bar{B}_{s}}$ mixing this
model can produce a new CP phase through contributions associated with
the operators $ B_{LL}= \bar{b} ( 1-\gamma_5) s \bar{b} (1- \gamma_5)
s $ and $ B_{RR}= \bar{b} ( 1+\gamma_5) s
\bar{b} (1+ \gamma_5) s $.  In the vacuum insertion approximation, the
matrix element of $B_{LL}$ and $B_{RR}$ are the same as only the
pseudoscalar currents contribute.  The contribution to $B_s$ mixing
comes from the sum 
$<S_{sb}^{*2} e^{ -i 2 \phi}B_{LL} +S_{bs}^2 e^{ i 2 \phi} B_{RR}> $
which
contain a term $ \sim i \chi_s \chi_a \sin { 2 \phi}$ which 
 is a source of new weak phase in
$B_{s}-{\bar{B}_{s}}$ mixing. This term will be small or vanish if
$\chi_a/\chi_s=\epsilon/z \sim 0$. Hence the presence of a new weak
phase in the $B_{s}-{\bar{B}_{s}}$ mixing requires the breaking of the
$s-b$ symmetry in the Yukawa coupling of the second Higgs doublet.

Finally, we note that the model will produce new effects in $b \to s
l^+l^-$ and $B_s \to l^+l^-$ decays which will depend on the couplings $s$
and $s_l$ of the second Higgs doublet to the quarks and leptons. A
detailed study of such processes in the model will be discussed
elsewhere \cite{futurework}.

\section {Up Quark Sector}
So far we have neglected the up sector and it is possible that there
will be FCNC decays in that sector also. However the diagonalizing of
the up quark mass matrix is connected to the CKM matrix via $V_{CKM}=
V_L^{\dagger} W$ where $W$ in Eq.~\ref{wmatrix} diagonalizes the down
quark matrix and $V^{\dagger}_L$ transform the left handed up type
quarks from the gauge to the mass basis. If we assume the up sector
has the same symmetry as the down sector then $V_L \sim W$. We can
write,
\bea
 W_u & = & 
 \pmatrix{1 & 0 &
0 \cr
0
& -\frac{1}{\sqrt{2}}(1-\frac{1}{2}z_u) & \frac{1}{\sqrt{2}}(1+\frac{1}{2}z_u)\cr
0
&\frac{1}{\sqrt{2}}(1+\frac{1}{2}z_u) &
 \frac{1}{\sqrt{2}}(1-\frac{1}{2}z_u)},\nonumber\\
 W_d & = & 
 \pmatrix{1 & 0 &
0 \cr
0
& -\frac{1}{\sqrt{2}}(1-\frac{1}{2}z_d) & \frac{1}{\sqrt{2}}(1+\frac{1}{2}z_d)\cr
0
&\frac{1}{\sqrt{2}}(1+\frac{1}{2}z_d) &
 \frac{1}{\sqrt{2}}(1-\frac{1}{2}z_d)},\
 \label{wmatrixupdown}
 \eea
 where   $z_u  \approx  \pm 2m_c/m_t$ and $z_d  \approx  \pm 2m_s/m_b$ break the 2-3 symmetry
 in the up and the down sector.
 The CKM matrix is now obtained as,
 \bea
 V_{CKM}&=& W_u^T \cdot W_d =
 \pmatrix{1 & 0 &
0 \cr
0
& 1+ \frac{z_uz_d}{4} & \frac{1}{2}(z_u-z_d)\cr
0
&\frac{1}{2}(z_d-z_u) &
1+ \frac{z_uz_d}{4} }.
\label{ckm}
\eea
We see that we get the right order and sign for the CKM element $V_{cb}$ and $V_{ts}$.
To obtain the realistic $V_{CKM}$ we have to introduce the Cabibbo angle $\lambda$
as a symmetry breaking effect.
 Possible 
 FCNC effects in the top sector will be discussed in a later work\cite{futurework}.

\section {Leptonic Sector}
In this section we study the consequences of  the $2-3$ symmetry  applied to the charged
lepton sector.
We  assume that the structure for the charged lepton mass
 matrix, $Y^L$ is given by,
\begin{eqnarray}
 Y^L &= &\pmatrix{l_{11} & 0 &
0 \cr
0
& l_{22}(1+2z_l) & l_{22}\cr
0
&l_{22} &
 l_{22}}, \
 \label{symbreaklepton}
 \eea
 where $z_l \sim 2m_{\mu}/m_{\tau}$ is a small number. This charged
 lepton mass matrix is now diagonalized by,
\bea
M^L_{diag} & = & {W_l }^{T} M^D W_l 
 =\pmatrix{ \pm \frac{v}{\sqrt{2}}l_{11} & 0 &
0 \cr
0
&  \pm \frac{v}{\sqrt{2}}z_l l_{22} & 0\cr
0
&0 &
 \pm \frac{v}{\sqrt{2}}(2+z_l)l_{22}}, \nonumber\\
 W_l & = & 
 \pmatrix{1 & 0 &
0 \cr
0
& -\frac{1}{\sqrt{2}}(1-\frac{1}{2}z_l) & \frac{1}{\sqrt{2}}(1+\frac{1}{2}z_l)\cr
0
&\frac{1}{\sqrt{2}}(1+\frac{1}{2}z_l) &
 \frac{1}{\sqrt{2}}(1-\frac{1}{2}z_l)}.\
 \label{wmatrixlepton}
 \eea
Just like the down quark sector we have broken the $\mu - \tau$ symmetry
by the $\mu$ mass. In the symmetry limit,
\bea
W_l  & \rightarrow &  U_l= \pmatrix{1 & 0 &
0 \cr
0
& -\frac{1}{\sqrt{2}} & \frac{1}{\sqrt{2}}\cr
0
&\frac{1}{\sqrt{2}} &
 \frac{1}{\sqrt{2}}}.\ 
\label{ullepton}
\eea
The Yukawa interaction associated with the second Higgs doublet is now
taken similar to Eq.~\ref{sdmsimple} as,
\bea
 S^L &= &e^{i \phi_l}\pmatrix{
 s_l & 0 &
0 \cr
0
& 0 &  \pm s_l( 1+ 2\epsilon_l)\cr
0
& \pm s_l & 0
}, \
\label{sdleptonsimple}
\eea
where $s_l$ and $\epsilon_l$ are real numbers and we have introduced a 
universal weak phase $\phi_l$. 
On moving to the mass eigenstate we obtain,
 \bea
S^L \rightarrow S^{L'} &= & e^{ i \phi_l}\pmatrix{s_l & 0 &
0 \cr
0
& \mp s_l( 1+ \epsilon_l)  & \pm s_l (z_l- \epsilon_l)\cr
0
&\pm s_l (z_l+ \epsilon_l) &
 \pm s_l(1+\epsilon_l) }. \
 \label{sdmleptonsimple}
\end{eqnarray}
To simplify our discussion  we  assume $\epsilon_l \sim 0$.
The model now   generates FCNC interactions of the $\tau$ like
$ \tau \to \mu \bar{l} l$ where $l$ are muon or electrons. Note that decays
$ \tau \to \mu \bar{l_a} l_b$ where $l_a \ne l_b$ are forbidden.
The effective Hamiltonian for such decays is easily written down as,
\bea
H_{l}^{eff} & = & \eta_{ll}\frac{1}{4}\frac{m_{\mu}}{m_{\tau}} \frac{s_l^2}{ m_H^2}
\left[\pm O_{LL}  \pm O_{RR} +O_{LR} +O_{RL} \right], \nonumber\\
O_{LL} & = & e^{ -i 2 \phi_l} \bar{\mu} ( 1-\gamma_5) \tau \bar{l} (1- \gamma_5) l, \nonumber\\
O_{RR} & = & e^{ i 2 \phi_l} \bar{\mu} ( 1+\gamma_5) \tau \bar{l} (1+ \gamma_5) l, \nonumber\\
O_{LR} & = &  \bar{\mu} ( 1-\gamma_5) \tau \bar{l} (1+ \gamma_5) l, \nonumber\\
O_{RL} & = &  \bar{\mu} ( 1+\gamma_5) \tau \bar{l} (1- \gamma_5) l. \
\label{hefflep}
\eea
To  further simplify the discussion we will 
choose $ \phi_l=0$. In ref.~\cite{he} the FCNC leptonic transitions
were studied in an effective operator formalism where the coefficient of the four quark operators were
taken to be $ \sim 4 \pi$. This lead to a constraint on the scale of NP, $\Lambda \sim 10$ TeV. 
To compare to our model we have the correspondence,
\bea
\frac{4 \pi}{\Lambda^2} & \equiv & \frac{1}{4}\frac{m_{\mu}}{m_{\tau}} \frac{s_l^2}{ m_H^2}.\
\eea
Choosing $ s_l \sim 1$ we find $m_H \sim 340 $ GeV. From $B$ decays we found
$m_{H} \sim 600 - 900$ GeV and so we  predict the branching ratio of
$ \tau \to \mu \bar{l} l$ to be below the experimental bounds \cite{pdg} by about
$(340/900)^4 - (340/600)^4 \sim 2 \times 10^{-2} - 10^{-1}$. A careful analysis would include varying the parameters
$ \phi_l$ and $s_l$ and will be presented elsewhere\cite{futurework}.
We can also study the decay 
$ \tau \to \mu \bar{q} q$ which can be studied in the  decays
$ \tau \to \mu \pi^0( \rho)$. The experimental bounds produce similar size limit on the Higgs mass as
 the decays $ \tau \to \mu \bar{l} l$. Note that for $ \phi_l=0$
 only $H_2$, which is a pseudoscalar, can contribute to
 $ \tau \to \mu \pi^0$ and the contribution from $H_1$ exchange vanishes.

Let us now turn to the neutrino sector.
The neutrino mixing, the PMNS matrix, arise
from the lepton mass Lagrangian as follows:
\begin{eqnarray}
{\cal L}_m~=~\nu^T_\alpha C^{-1}{\cal M}_{\nu,\alpha\beta}\nu +
\bar{e}_{\alpha,L}M^e_{\alpha \beta}e_R + h.c.
\end{eqnarray}
Diagonalizing the mass matrices by the transformations
$U^T_\nu{\cal M}_\nu U_\nu={\cal M}^\nu_{diag}$ and $U^{\dagger}_\ell
M^eV~ =M^e_{diag}$, one defines the neutrino mixing matrix as
$U_{PMNS}~=~U^{\dagger}_\ell U_\nu$. 
We will parameterize $U_{PMNS}$ as follows:
\begin{eqnarray}
 U_{PMNS}=\pmatrix{c_{12}c_{13} & s_{12}c_{13} &
s_{13} e^{-i\delta} \cr
-s_{12}c_{23}-c_{12}s_{23}s_{13}e^{i\delta}
&c_{12}c_{23}-s_{12}s_{23}s_{13}e^{i\delta} & s_{23}c_{13}\cr
s_{12}s_{23}-c_{12}c_{23}s_{13}e^{i\delta}
&-c_{12}s_{23}-c_{12}c_{23}s_{13}e^{i\delta} &
 c_{23}c_{13}}K,
\end{eqnarray}
where $K~=~diag(1, e^{i\phi_1},e^{i\phi_2})$.

The 2-3 symmetry in the leptonic sector leads to a simplified form of
the PMNS matrix, with $s_{13}=0$, 
given by,
\bea
U_{PMNS}^{s} & = & \pmatrix{c_{12} &s_{12}&0\cr
-\rr2 s_{12}
&\rr2 c_{12}  & \rr2\cr
\rr2 s_{12}
&- \rr2 c_{12} &
 \rr2}.
\label{usym}
\eea
In the basis where the mass matrix of the charged leptons is diagonal,
the left-handed flavor-based (symmetric)
neutrino mass matrix $m'$ is related to its diagonal form,
$M_{\nu,diag}=diag[m_1,m_2,m_3]$ by, 
\bea
U^T_{PMNS}M_{\nu}^{\prime}U_{PMNS}=M_{\nu,diag}.
\label{numass}
\eea
The form of $U_{PMNS}^{s}$
 just follows from the $ \mu-\tau $ symmetry in
$M_{\nu}^{\prime}$ \cite{mutau}.

Note that we can express $U_{PMNS}^{s}$, using Eq.~\ref{ullepton}, as,
\bea
U_{PMNS}^{s}&= U^{\dagger}_\ell U_\nu, \
\label{nulepton}
\eea
where
\bea
U^{\dagger}_\ell & = &
\pmatrix{1 & 0 &
0 \cr
0
& -\frac{1}{\sqrt{2}} & \frac{1}{\sqrt{2}}\cr
0
&\frac{1}{\sqrt{2}} &
 \frac{1}{\sqrt{2}}},\nonumber\\
U_\nu & = & \pmatrix{c_1&-s_1&0\cr s_1&c_1&0\cr 0&0&1\cr}\pmatrix{1&0&0\cr 0& -1&0\cr 0&0&1\cr}.\ 
\eea 
So the neutrino matrix, $U_{\nu}$ is just a combination of a simple rotation matrix and a 
phase matrix. 
Now the breaking of the 2-3 symmetry in the leptonic sector will
change $U^{\dagger}_\ell$ and $U_{PMNS}$   to,
\bea
U^{\dagger}_\ell & \rightarrow & W_l = \pmatrix{1 & 0 &
0 \cr
0
& -\frac{1}{\sqrt{2}}(1-\frac{1}{2}z_l) & \frac{1}{\sqrt{2}}(1+\frac{1}{2}z_l)\cr
0
&\frac{1}{\sqrt{2}}(1+\frac{1}{2}z_l) &
 \frac{1}{\sqrt{2}}(1-\frac{1}{2}z_l)},\nonumber\\
U_{PMNS} & \rightarrow & \pmatrix{c_{12} &s_{12}&0\cr
-\rr2 s_{12}(1-\frac{z_l}{2})
&\rr2 c_{12}(1-\frac{z_l}{2})  & \rr2(1+\frac{z_l}{2})\cr
\rr2(1+\frac{z_l}{2}) s_{12}
&- \rr2 c_{12}(1+\frac{z_l}{2}) &
 \rr2(1-\frac{z_l}{2})},\
 \label{theta23}
 \eea
 where  $z_l= \pm 2{m_{\mu} \over m_{\tau}}$.
 This corresponds to $ s_{23}=\rr2(1+\frac{z_l}{2})$ and $c_{23}=\rr2(1-\frac{z_l}{2})$ and to 
 an atmospheric mixing angle of $ \theta_{23} \sim 43.26^0$ for the negative sign of $z_l$.

 Finally, we point out that this model will have interesting collider
 signatures. The heavy Higgs, $H$, according to the structure in
 Eq.~\ref{sdmsimple}, couples equally to all three generations, 
to a very good approximation, unlike
 the usual SM Higgs which has couplings proportional to the mass.  
The effect of electroweak
 precision measurements and collider signatures will be explored
 elsewhere\cite{futurework}.
 
\section{Conclusions}
 In summary, there are now several $B$ decay modes in which there
 appear to be deviations from the SM predictions. These deviations
 could signal the presence of beyond the SM physics.  In this work we
 were interested in a NP scenario that can provide a single solution
 to all the deviations.  We considered a two Higgs doublet model with
 a 2-3 symmetry in the down type quark and the charged lepton sector.
 The breaking of the 2-3 symmetry, introduced by the strange quark
 mass and the muon mass lead to FCNC in the quark sector and the
 charged lepton sector that are suppressed by ${ m_s \over m_b}$ and
 ${ m_{\mu} \over m_{\tau}}$ in addition to the mass of the heavy
 Higgs boson.  Additional FCNC effects of similar size were generated
 from the breaking of the $s-b$ symmetry in the Yukawa coupling of the
 second Higgs doublet.  {} From a fit to the $ B \to K \pi $ data we
 found the mass of the lightest neutral Higgs in the second doublet to be of
 the order $ m_H \sim 600 - 900$ GeV. We made several predictions in $B$
 decays listed below:
 
\begin{itemize}

\item{  A sizeable transverse polarization in the decays $\bdbar \to \phi K^{*}$ and
$B^- \to \rho^- K^{0*}$ was predicted but not in $B^- \to \rho^0
K^{*-}$. This is consistent with present measurements \cite{rhoKstarexp,hfag}.
We also predicted the possibility of observing a
    sizeable triple product asymmetry and/or direct CP asymmetry in
    the decays $\bdbar \to \phi K^{*}$ and
$B^- \to \rho^- K^{0*}$.  
}

\item{  The $\sin { 2 \beta}$ measurement 
in $ \bd(t) \to \phi K_s$ was predicted to show a small deviation from the SM 
value( if Eq.~\ref{sdmsimple} is assumed) but it was found that 
$\sin { 2 \beta}$ measured in 
$ \bd(t) \to \eta^{\prime} K_s$ could have significant deviation from the
SM prediction. 
}

\item{ The decays with the quark transition $b \to s \bar{u}u$ and $b \to s \bar{c}c$ were found to be unaffected
}

\item{We found a new source of weak phase from our model
 in $B_{s}-{\bar{B}_{s}}$ mixing 
while
$B_{d}-{\bar{B}_{d}}$ mixing was not affected.
}  
\end{itemize}
We studied the implication of the 2-3 symmetry extended to the up
sector and found that we could generate the right sign and size of the
CKM matrix element $V_{cb}$ and $V_{ts}$.  We then studied FCNC
effects in the lepton sector.  The lepton flavour violating decays $
\tau \to \mu \bar{l}(\bar{q}) l(q)$ were found to be below 
the present experimental
limits by a factor of $ 2 \times 10^{-2} - 10^{-1}$. 
Finally, we found that the breaking of 2-3 symmetry in the
lepton sector could lead to deviations of the atmospheric neutrino
mixing angle from the maximal value by $ \sim 2$ degrees.

\bigskip
\noindent {\bf Acknowledgements}:
This work is financially supported by NSERC of Canada. We thank David London, R.N. Mohapatra and Sandip Pakvasa for useful discussion.



\begin{thebibliography}{99}

\bibitem{mutau} C. S. Lam, hep-ph/0104116; T. Kitabayashi and
M. Yasue, Phys.Rev. {\bf D67} 015006 (2003); W. Grimus and L. Lavoura,
hep-ph/0305046; 0309050; Y. Koide,  Phys.Rev. {\bf D69}, 093001 (2004);
for examples of such theories, see W. Grimus and
L. Lavoura, hep-ph/0305046; 0309050.

\bibitem{moh}  R. N. Mohapatra, SLAC Summer
Inst. lecture; http://www-conf.slac.stanford.edu/ssi/2004; hep-ph/0408187;
JHEP, {\bf 0410}, 027 (2004);
  W. Grimus, A. S.Joshipura, S. Kaneko, L.
Lavoura, H. Sawanaka, M. Tanimoto, hep-ph/0408123; A.~Ghosal,
  Mod.\ Phys.\ Lett.\ A {\bf 19}, 2579 (2004).


\bibitem{Babar}
B.~Aubert {\it et al.}  [BABAR Collaboration],
  arXiv:hep-ex/0507087;
B.~Aubert {\it et al.}  [BABAR Collaboration],
[hep-ex/0503011]; B.~Aubert {\it et al.}  [BABAR Collaboration],
[hep-ex/0502019]; B.~Aubert {\it et al.}  [BABAR Collaboration],
[hep-ex/0502017];
B.~Aubert {\it et al.}  [BABAR Collaboration],
  Phys.\ Rev.\ Lett.\  {\bf 94}, 161803 (2005)
  [arXiv:hep-ex/0408127].

\bibitem{Belle}
K.~Abe {\it et al.}  [Belle Collaboration],
  arXiv:hep-ex/0507037;
K.~F.~Chen {\it et al.}  [Belle Collaboration],
  Phys.\ Rev.\ D {\bf 72}, 012004 (2005)
  [arXiv:hep-ex/0504023];
K.~Abe {\it et al.}  [BELLE Collaboration],
  arXiv:hep-ex/0409049.
  
 

\bibitem{hfag}
{Hevay flavor average
  group,http://www.slac.stanford.edu/xorg/hfag/rare}.




\bibitem{BKpidecays1} Experiment: B.~Aubert {\it et al.}  [BABAR
Collaboration], Phys.\ Rev.\ Lett.\ {\bf 89}, 281802 (2002);
arXiv:hep-ex/0408062, arXiv:hep-ex/0408080, arXiv:hep-ex/0408081;
A.~Bornheim {\it et al.}  [CLEO Collaboration], Phys.\ Rev.\ D {\bf
68}, 052002 (2003); Y.~Chao {\it et al.}  [Belle Collaboration],
Phys.\ Rev.\ D {\bf 69}, 111102 (2004). Theory: A.~J.~Buras,
R.~Fleischer, S.~Recksiegel and F.~Schwab, Eur.\ Phys.\ J.\ C {\bf
32}, 45 (2003), Phys.\ Rev.\ Lett.\ {\bf 92}, 101804 (2004), Nucl.\
Phys.\ B {\bf 697}, 133 (2004), arXiv:hep-ph/0410407; V.~Barger,
C.~W.~Chiang, P.~Langacker and H.~S.~Lee, Phys.\ Lett.\ B {\bf 598},
218 (2004); S.~Mishima and T.~Yoshikawa, arXiv:hep-ph/0408090;
Y.~L.~Wu and Y.~F.~Zhou, arXiv:hep-ph/0409221; H.~Y.~Cheng, C.~K.~Chua
and A.~Soni, arXiv:hep-ph/0409317; Y.~Y.~Charng and H.~n.~Li,
arXiv:hep-ph/0410005; X.~G.~He and B.~H.~J.~McKellar,
arXiv:hep-ph/0410098;
C.~S.~Kim, S.~Oh and C.~Yu,
  arXiv:hep-ph/0505060.

\bibitem{BKpidecays2a}  
S.~Baek, P.~Hamel, D.~London,
A.~Datta and D.~A.~Suprun, Phys.\ Rev.\ D {\bf 71}, 057502 (2005)
[arXiv:hep-ph/0412086].




\bibitem{dattaKX}
  T.~E.~Browder, A.~Datta, X.~G.~He and S.~Pakvasa,
  Phys.\ Rev.\ D {\bf 57}, 6829 (1998)
  [arXiv:hep-ph/9705320].




\bibitem{BVVTP} For a study of triple products in the SM and with new
physics, see A.~Datta and D.~London, Int.\ J.\ Mod.\ Phys.\ A {\bf
19}, 2505 (2004).

\bibitem{BaBarTP} B.~Aubert {\it et al.} [BABAR Collaboration],
arXiv:hep-ex/0408017. Note that the earlier Belle measurements of the
same quantities do not show any signs of a nonzero triple-product
asymmetry, see K.-F. Chen {\it et al.}  [Belle Collaboration], Phys.\
Rev.\ Lett.\ {\bf 91}, 201801 (2003).

\bibitem{phiKstarexp} B.~Aubert {\it et al.}  [BABAR Collaboration],
Phys.\ Rev.\ Lett.\ {\bf 91}, 171802 (2003); K.~F.~Chen {\it et al.}
[BELLE Collaboration], arXiv:hep-ex/0503013.
 

\bibitem{rhoKstarexp} Particle Data Group, Ref.~\cite{pdg};
B.~Aubert [BABAR Collaboration], arXiv:hep-ex/0408093; K.~Senyo [Belle
Collaboration].

\bibitem{rhoKstar}  
S.~Baek, A.~Datta, P.~Hamel, O.~F.~Hernandez and D.~London,
  arXiv:hep-ph/0508149.


\bibitem{susygut} See for e.g.  T.\ Moroi, JHEP.\ {\bf
0003} 019 (2000),
  [arXiv:hep-ph/0002208];  N.\ Akama, Y.\ Kiyo, S.\ Komine and T.\ Moroi,
Phys.\ Rev.\ {\bf D64} 095012 (2001),
  [arXiv:hep-ph/0104263];
  D.~Chang, A.~Masiero and H.~Murayama,
  Phys.\ Rev.\ D {\bf 67}, 075013 (2003)
  [arXiv:hep-ph/0205111] and references therein.

\bibitem{BBNS} M. Beneke, G. Buchalla, M. Neubert and
C.T. Sachrajda, Nucl.\ Phys.\ B {\bf 591}, 313 (2000), Nucl.\ Phys.\ B
{\bf 606}, 245 (2001).

\bibitem{BKpidecays2b}  
 A.~Datta, M.~Imbeault, D.~London, V.~Page,
N.~Sinha and R.~Sinha, Phys.\ Rev.\ D {\bf 71}, 096002 (2005);
A.~Datta and D.~London,
  Phys.\ Lett.\ B {\bf 595}, 453 (2004)
  [arXiv:hep-ph/0404130].

\bibitem{futurework} Alakabha Datta, work in progress.

\bibitem{soni}
  D.~Atwood, L.~Reina and A.~Soni,
  Phys.\ Rev.\ D {\bf 55}, 3156 (1997)
  [arXiv:hep-ph/9609279].


\bibitem{Das}
  A.~K.~Das and C.~Kao,
  Phys.\ Lett.\ B {\bf 372}, 106 (1996)
  [arXiv:hep-ph/9511329].
  


  
\bibitem{pdg} S.~Eidelman {\it et al.}  [Particle Data Group
Collaboration], Phys.\ Lett.\ B {\bf 592} (2004) 1,
http://pdg.lbl.gov/pdg.html.

\bibitem{dattarparity} A.~Datta,
  Phys.\ Rev.\ D {\bf 66}, 071702 (2002)
  [arXiv:hep-ph/0208016].





\bibitem{dattaeta}
  A.~Datta, X.~G.~He and S.~Pakvasa,
  Phys.\ Lett.\ B {\bf 419}, 369 (1998)
  [arXiv:hep-ph/9707259].

\bibitem{lipkin}H.~J.~Lipkin,
Phys.\ Lett.\ B {\bf 254}, 247 (1991),   Phys.\ Lett.\ B {\bf 544}, 145 (2002),
Phys.\ Lett.\ B {\bf 283}, 421 (1992),
A.~Datta, H.~J.~Lipkin and P.~J.~O'Donnell,
  Phys.\ Lett.\ B {\bf 544}, 145 (2002)
  [arXiv:hep-ph/0206155];
H.~J.~Lipkin,
  arXiv:hep-ph/0507225.

\bibitem{gammaDD}
R.~Fleischer,
  Eur.\ Phys.\ J.\ C {\bf 10}, 299 (1999);
  [arXiv:hep-ph/9903455].
 A.~Datta and D.~London, JHEP {\bf 0404}, 072
(2004); A.~Datta and D.~London, Phys.\ Lett.\ B {\bf 584}, 81
(2004); J.~Albert, A.~Datta and D.~London,
 Phys.\ Lett.\ B {\bf 605}, 335 (2005)
  [arXiv:hep-ph/0410015].



\bibitem{he}
  D.~Black, T.~Han, H.~J.~He and M.~Sher,
  Phys.\ Rev.\ D {\bf 66}, 053002 (2002)
  [arXiv:hep-ph/0206056].

\end{thebibliography}
\end{document}